\documentclass[aps,pre,twocolumn,showpacs]{revtex4-1}
\usepackage{amsmath,amsfonts,amssymb,wrapfig,lipsum,color,soul,hyperref}

\DeclareFontFamily{U}{rsfs}{}         % Formal Script            %
\DeclareFontShape{U}{rsfs}{m}{n}{<5> rsfs5 <6><7> rsfs7          %
  <8><9><10><10.95><12><14.4><17.28><20.74><24.88> rsfs10}{}     %
\DeclareMathAlphabet{\mathfs}{U}{rsfs}{m}{n}                     %
\newcommand{\mfs}[1]{\mathfs {#1}}                               %
\sethlcolor{red}
\definecolor{indiagreen}{rgb}{0.07, 0.53, 0.03}
\def\beq{\begin{eqnarray}}
\def\eeq{\end{eqnarray}}

\def\nn{\nonumber\\}

\def\lie{\mfs L}

\def\half{{\textstyle{\frac{1}{2}}}}
\begin{document}
\title{Note on Kerr/CFT correspondence in a first order formalism}

\author{Avirup Ghosh}\email{avirup.avi@gmail.com}
\affiliation{Saha Institute of Nuclear Physics, 1/AF Bidhan Nagar, Kolkata 700064, India.}

\begin{abstract}
In symmetry based approaches to black hole entropy, we calculate the central charge of the Virasoro algebra in the first order formulation of gravity for both Palatini and Holst actions. In these calculations, we made use of the NHEK metric and the Kerr-CFT correspondence. For the Palatini action the results obtained in the second order formulation are reproduced. We also argue that the Holst term does not contribute to the charge algebra no matter what geometry/boundary conditions one is considering.
\end{abstract}
\maketitle

\section{Introduction}
The symmetry based approaches or the dual holographic description of the black hole entropy has its origin in the work of Brown and Henneaux\cite{Brown:1986nw}. The essential point is to argue that the quantum theory of black hole should be a  holographic dual of a 2d conformal field theory (CFT) living at the spacetime boundary. One then expects that the states in some unitary representation of the CFT are the microstates of the black hole. The actual construction of the CFT is still elusive. So far, one has been able to only count the microstates rather than explicitly construct a dual CFT. Even the location of the boundary is debatable. In \cite{Brown:1986nw}, the boundary is taken at asymptotic infinity, while in some other calculations the boundary is taken as the black hole horizon (see \cite{symmetry} and references therein). The main idea behind this approach is to identify the 2d conformal symmetry group, isomorphic to $Diff(S^1)$, that is expected to be the symmetry group of the holographic quantum 
theory of black holes. The states of this quantum theory, which are 
possibly the black hole microstates in question, would then furnish a representation for this symmetry group. The representation is expected to be characterised by the appropriate black hole parameters, such as the horizon area, charges and angular momentum. The black hole entropy would then be equal to the logarithm of the dimension of such a representation (for example, the number of quantum states for a fixed area, angular momentum and charges). From the outset, the aim is to calculate only the number of micro-states knowing that from the symmetry group alone it would be impossible to label a complete set of microstates. When one looks for symmetries near the horizon or at asymptotic infinity, the usual notion of symmetries represented by exact Killing vectors is not enough. One, therefore, uses an extended notion of symmetries in terms of approximate Killing vectors that gives a larger set of vector fields. The set of such approximate Killing fields are determined by certain fall-off conditions on the 
metric.

The Kerr-CFT correspondence, originally initiated by \cite{Guica:2008mu} is parallel to the idea of Brown and Henneaux except for the fact that the background spacetime is now the Near horizon Extremal Kerr (NHEK), which is topologically  $AdS_2\times S^2$, rather than $AdS_3$ as in the case of Brown and Henneaux. The appropriate boundary in NHEK is a timelike boundary.

In symmetry based approaches the issue of the correct Poisson brackets of charges (in the second order formulation) is not completely resolved and is tied to the choice of boundary conditions. There exist more than one ways of calculating the Poisson brackets of charges \cite{{Barnich:2001jy},{silva},{Majhi:2011ws}} and it seems that their algebra having a central extension in one calculation may have different or no central extension at all in some other calculations. On the other hand the symplectic structure in the first order formulation is clean and studied in detail in the context of asymptotic symmetries \cite{{Corichi:2010ur},{Ashtekar:2008jw}} and laws of black hole mechanics \cite{{Ashtekar:1999wa},{Ashtekar:1998sp},{Ashtekar:1999yj},{Chatterjee:2006vy},{Chatterjee:2008if}}. Therefore, it seems justified to apply the first order symplectic structure to a well-studied case---the Kerr-CFT correspondence. 

In the first order formulation, apart from studying the bulk symplectic structure one also studies the boundary symplectic structure. Depending on the boundary conditions, one may need to add a boundary symplectic current to avoid leakage of any flux across the boundary. This ensures that the symplectic structure is hypersurface independent. This subtle issue is apparently overlooked in the existing calculations of Poisson brackets of charges in the second order formulation. This has already been pointed out in \cite{Amsel:2009pu} for instance.

We also study the effect of adding the Holst term to the action. It is already known \cite{Rovelli:1997na} that in presence of the Holst term different values of the Immirzi parameter yield nonequivalent quantum gravity theories and a particular choice is to be made to recover the Bekenstein-Hawking(BH) entropy from the exact counting of states in the quantum theory \cite{Ashtekar:1997yu}. An intriguing question is does the semiclassical symmetry based approach retain any imprint of the Immirzi parameter? We argue that if one works with a hypersurface independent symplectic structure then the Holst term will never contribute, no matter what geometry/boundary conditions one is considering. This is in agreement with some recent calculations \cite{Ghosh:2011fc} where it is claimed that the semiclassical limit of black hole entropy in LQG does not depend on particular choices of the Immirzi parameter.

In the context of Wald entropy the Holst term in presence of a negative cosmological constant has been studied in \cite{Durka:2011yv}. It has been shown that the Immirzi parameter does not play a role for AdS-Schwarzschild and AdS-Kerr spacetimes but makes a nontrivial contribution to the entropy and mass for AdS-Taub-Nut spacetime. (Similar results have been obtained employing Euclidean path integrals in \cite{Liko:2011cq}). In recent past some attempts have been made to compare the Wald entropy and the entropy from symmetry based approaches from some alternative construction of the Poisson bracket algebra \cite{Majhi:2011ws}. Our results show that Wald entropy and entropy from symmetry based approaches might not always match.

The Kerr-CFT correspondence has been generalised to an Isolated-Horizon CFT correspondence in \cite{Wu:2009di}. In this case the metric in the neighbourhood of an axisymmetric extremal isolated horizon has been used and a calculation similar to the one in the Kerr-CFT correspondence has been carried out. However, a study of the `near-horizon' symmetries of an isolated horizon is still missing. Since isolated horizons are studied primarily in the first order formulation, our exercise might shed some light into a symmetry based approach to isolated horizons. 

In this note, we start with the NHEK metric and redo the calculations of the Poisson brackets of charges in the first order formulation of gravity (all of the calculations that have appeared till now has been in the second order formulation). We also study the effect of adding Holst term to the action. This gives some insight into what role the Holst term plays in the semi-classical regime.

%%%%%%%%%%%%%%%%%%%%%%%%%%%%%%%%%%%%%%%%%%%%%%%%%%%%%%%%%%%%%%%%%%%%%%%%%%%%%%%%
%%%%%%%%%%%%%%%%%%%%%%%%%%%%%%%%%%%%%%%%%%%%%%%%%%%%%%%%%%%%%%%%%%%%%%%%%%%%%%%%
%%%%%%%%%%%%%%%%%%%%%%%%%%%%%%%%%%%%%%%%%%%%%%%%%%%%%%%%%%%%%%%%%%%%%%%%%%%%%%%%
%%%%%%%%%%%%%%%%%%%%%%%%%%%%%%%%%%%%%%%%%%%%%%%%%%%%%%%%%%%%%%%%%%%%%%%

\section{The NHEK metric and boundary conditions}

\subsection{Boundary conditions}
The NHEK geometry which has an $SL(2,R)\times U(1)$ isometry group has been studied in detail in \cite{Bardeen:1999px}. We would not go into the details of the NHEK geometry except for the fact that in some global 
coordinate system the NHEK metric takes the form:
\begin{align}
ds^2=2GJ\Omega^2\Big(-(1&+r^2)dt^2+\frac{dr^2}{1+r^2}+\nn
&+d\theta^2+\Lambda^2(d\phi^2+rdt^2)\Big)
\end{align}

The tetrads can then be obtained such that they staisfy $g_{\mu\nu}=\eta_{IJ}e_\mu^Ie_\nu^J$, where $\eta_{IJ}$ is the Minkowski metric. It then follows that the tetrads can be taken to be,
\beq\label{backT}
~~~~~~~~~~~&&e^0=N\sqrt{1+r^2}dt~~~~~~~~e^1=\frac{Ndr}{\sqrt{1+r^2}}\nn
\nn
&&e^2=Nd\theta~~~~~~~~~~~~~~~e^3=N \Lambda(d\phi+rdt)\nn
\\
&&N=(2JG\Omega^2)^{\frac{1}{2}}~~~~~~
\Omega^2=\frac{1+cos^2\theta}{2}\nn
&&~~~~~~~~~~~~~\Lambda=\frac{2sin\theta}{1+cos^2\theta}
\nn
\eeq
The range of the coordinates are $0\leq\theta<\pi$ and $0\leq\phi<2\pi$ and the boundary at $r\rightarrow\infty$ is a time-like boundary.

\hspace{2cm}

The connection can be calculated from the torsion free condition, $de^I+\omega^I~_J\wedge~e^J=0$. It can be recast in the form,
\beq\label{backConn}
\omega^{IJ}_\mu=e^{I\nu}\nabla_{\mu}e^{J}_{\nu}
\eeq
where $\nabla_\mu$ is the usual covariant derivative compatible with the metric. It then follows that,
\beq
\omega^{10}&=&\frac{1}{2}r(\Lambda^2-2)dt+\frac{1}{2}\Lambda^2d\phi~~~~~
\omega^{20}=-\frac{\sqrt{1+r^2}\frac{dN}{d\theta}}{N}dt\nn
\nn
\omega^{30}&=&\frac{1}{2}\frac{\Lambda}{\sqrt{1+r^2}}dr~~~~~~~~~~~~~~~~~~
\omega^{21}=-\frac{\frac{dN}{d\theta}}{N\sqrt{1+r^2}}dr\nn
\nn
\omega^{31}&=&\frac{1}{2}\Lambda\sqrt{1+r^2}dt\nn
\nn
\omega^{32}&=&\frac{r}{N}\frac{d(N\Lambda)}{d\theta}dt+\frac{1}{N}\frac{d(N\Lambda)}{d\theta}d\phi\nn
\eeq
Under the boundary conditions assumed in \cite{Guica:2008mu}:

\beq\label{strictbc}
\left(
  \begin{array}{ccccc}
 h_{tt}= O({r^2}) & h_{t\phi}= O({1}) & 
h_{t\theta}= O({1\over r}) &h_{t r}= O({1\over r^2})  
\\
 h_{\phi t}=h_{t\phi} & h_{\phi\phi}= O(1) &h_{\phi\theta}= 
O({1\over r})  &h_{\phi r}= O({1\over r})  \\
   h_{\theta t}=h_{t\theta} & h_{\theta\phi}=h_{\phi\theta} & 
h_{\theta\theta}= O({1\over r}) &h_{\theta r}= O({1\over 
r^2}) \\
   h_{rt}=h_{t r} & h_{r\phi}=h_{\phi r} & h_{r\theta}=h_{\theta r} 
& h_{rr}= O({1\over r^3}) \\
  \end{array}
\right)\ ,\nn \eeq
The asymptotic symmetry generating vector fields can be calculated using,
\beq
\lie_\xi g_{\mu\nu}=h_{\mu\nu}
\eeq
and then equating terms of same orders in $r$ on both sides. It then follows that symmetry generating vector fields are of the form,
\beq\label{allw} \xi_A = \left(-r
\epsilon'(\phi) + O({1})\right)\partial_r + \left(C+O\left({1\over r^3}\right)\right)\partial_t
 + \nn
 +\left(\epsilon(\phi) + O\left({1\over r^2}\right)\right)\partial_\phi + O\left({1\over r}\right)\partial_\theta\nn
\eeq
where the higher order terms generate trivial diffeomorphisms. The relevant subalgebra isomorphic to a $Diff(S^1)$ is then,
\beq
\xi=\epsilon\frac{\partial}{\partial\phi}- r\epsilon'\frac{\partial}{\partial r}
\eeq
with $\epsilon(\phi)=-e^{-im\phi}$.
%%%%%%%%%%%%%%%%%%%%%%%%%%%%%%%%%%%%%%%%%%%%%%%%%%%%%%%%%%%%%%%%%%%%%%%%%%%%%%%%
%%%%%%%%%%%%%%%%%%%%%%%%%%%%%%%%%%%%%%%%%%%%%%%%%%%%%%%%%%%%%%%%%%%%%%%%%%%%%%%%
%%%%%%%%%%%%%%%%%%%%%%%%%%%%%%%%%%%%%%%%%%%%%%%%%%%%%%%%%%%%%%%%%%%%%%%%%%%%%%%%
%%%%%%%%%%%%%%%%%%%%%%%%%%%%%%%%%%%%%%%%%%%%%%%%%%%%%%%%%%%%%%%%%%%%
\subsection{Asymptotic expansion of tetrads}
The tetrads  and the connection can be expanded in a power series.
\beq
e^I&=&^0e^I+~\frac{^1e^I}{r}+~\frac{^2e^I}{r^2}+...\nn
\omega^{IJ}&=&^0\omega^{IJ}+~\frac{^1\omega^{IJ}}{r}+~\frac{^2\omega^{IJ}}{r^2}+...
\eeq
Unlike the asymptotically flat case \cite{Ashtekar:2008jw} where $^0e^I$ is just the Minkowski tetrad and fixed in the phase space, here it does vary because of boundary conditions imposed. So rather than taking the ANHEK (from here on ANHEK, would mean the asymptotic form of the NHEK metric) tetrad as the zeroth order one, we take the following. 

\beq\label{GENTE}
^0e^0&=&NA(t,\theta,\phi)rdt~~~~~^0e^1=\frac{Ndr}{r}+NB(t,\theta,\phi)d\phi\nn
~~~~~^0e^2&=&Nd\theta~~~~~^0e^3=\frac{N \Lambda}{C(t,\theta,\phi)}d\phi+N\Lambda C(t,\theta,\phi)rdt\nn
\eeq
We retain the terms that go like $ rdt,\frac{dr}{r},d\phi,d\theta$ at the zeroth order. We assume certain regularity conditions to hold on $A,B,C$ to ensure that the tetrads don't become  degenerate for any values of $\theta$ and $\phi$.
We note that the asymptotic metric calculated with this is:

\beq
ds^2&=&2GJ\Omega^2\left(-(A(t,\theta,\phi)^2-\Lambda^2C(t,\theta,\phi)^2)r^2dt^2+\frac{dr^2}{r^2}\right.\nn
&&~~~~~~~~~~~~~+\left.2\frac{B(t,\theta,\phi)}{r}drd\phi
+d\theta^2+2r\Lambda^2 dtd\phi\right.\nn
&&~~~~~~~~~~~~~~~~~~+\left.\left(\frac{\Lambda^2}{C(t,\theta,\phi)^2}+B(t,\theta,\phi)^2\right)d\phi^2 \right)\nn
\eeq
which is in agreement with the fall-off conditions. Moreover with the replacement.
\beq\label{eq:infinitesimal}
C(t,\theta,\phi)=A(t,\theta,\phi)=1+\eta F(t,\phi)\nn
\nn
B(t,\theta,\phi)=\eta \partial_{\phi}F(t,\phi)
\eeq
correctly reproduces the asymptotic constraints \cite{Guica:2008mu} at linear order in $\eta$ and leading order in $r$. For completeness we spell out these conditions. For perturbations $h_{\mu\nu}$ about the NHEK metric the asymptotic contraints imply
\beq
h_{\phi\phi} &=& \Lambda^2 \Omega^2 f (t, r, \phi)\nn
\nn
h_{tt}&=&r^2(1-\Lambda^2)\Omega^2f (t, r, \phi)\nn
\nn
h_{r\phi}&=&-\frac{\Omega^2}{2r}\partial_\phi f (t, r, \phi)
\eeq

Any other contribution to the tetrad consistent with the boundary conditions enter $^1e^I$ and higher order terms in the asymptotic expansion. A typical form of $^1e^I$ would be:
\beq
^1e^0&=& A_1(t,\theta,\phi)rdt+A_2(t,\theta,\phi)d\phi\nn
\nn
^1e^1&=& B_1(t,\theta,\phi)\frac{dr}{r}+B_2(t,\theta,\phi)d\theta+B_3(t,\theta,\phi)d\phi\nn
\nn
^1e^2&=& C_1(t,\theta,\phi)\frac{dr}{r}+C_2(t,\theta,\phi)d\theta+C_3(t,\theta,\phi)d\phi\nn
\nn
^1e^3&=& D_1(t,\theta,\phi)rdt+D_2(t,\theta,\phi)d\phi\nn
\eeq
One can check that this in agreement with the boundary conditions.

%%%%%%%%%%%%%%%%%%%%%%%%%%%%%%%%%%%%%%%%%%%%%%%%%%%%%%%%%%%%%%%%%%%%%%%%%%%%%%%%
%%%%%%%%%%%%%%%%%%%%%%%%%%%%%%%%%%%%%%%%%%%%%%%%%%%%%%%%%%%%%%%%%%%%%%%%%%%%%%%%
%%%%%%%%%%%%%%%%%%%%%%%%%%%%%%%%%%%%%%%%%%%%%%%%%%%%%%%%%%%%%%%%%%%%%%%%%%%%%%%%
%%%%%%%%%%%%%%%%%%%%%%%%%%%%%%%%%%%%%%%%%%%%%%%%%%%%%%%%%%%%%%%%%%%%%%%%%%%%%%%%
%%%%%%%%%%%%%%%%%%%%

\section{Palatini Action}
\subsection{Symplectic Structure}
The Palatini action in first order gravity is given by:
\beq\label{lagrangian}
S=-\frac{1}{16\pi G}\int_{\mathcal{M}}\left(\Sigma_{I\!J}\wedge 
F^{I\!J}\right)\;
\eeq
where $\Sigma_{IJ}=\half\,\epsilon_{IJ}{}_{KL}e^K\wedge e^L$, $\omega^{IJ}$ is a Lorentz $SO(3,1)$ connection and $F^{IJ}$ is a curvature two-form corresponding to the connection given by
$F^{IJ}=d\omega^{IJ}+\omega^I~_K\wedge \omega^{KJ}$. The action might have to be supplemented with boundary terms to make the variation well defined. But that does not effect the symplectic structure $\Omega(\delta_1,\delta_2)$, since 
$\delta_1,\delta_2$ are independent variations(i.e. they commute).

\vspace{.2cm}
On-shell the variation of the Lagrangian gives $\delta L=d\Theta(\delta)$ where $16\pi G
\Theta(\delta)=-\Sigma_{I\!J}\wedge \delta \omega^{I\!J}$. One then constructs the symplectic structure $\Omega$ on the space of
solutions. One first constructs the symplectic current $J(\delta_1,\delta_2)=
\delta_1\Theta(\delta_2)-\delta_2\Theta(\delta_1)$, which is closed on-shell. The symplectic structure is then given by:
\beq
\Omega(\delta_1,\delta_2)=\int_{M}J(\delta_1,\delta_2)=-\frac{1}{8\pi 
G}\int_{M}\left(\delta_{[1}\Sigma_{IJ}\wedge\delta_{2]} \omega^{IJ}\right)\nn
\eeq
where $M$ is a Cauchy surface.

A point to note here is that there can be non-trivial contributions from the boundary symplectic structure. We consider the symplectic current 3-form for the Palatini action. It follows that on shell

\beq
dJ&=&0
\eeq
this implies that when integrated over a closed region of spacetime bounded by $M_1\cup M_2\cup B$ (where B is a portion of the boundary of spacetime given by $r\rightarrow \infty$ in our case),
\beq
\int_{M_1}J&-&\int_{M_2}J~+~\int_{r\rightarrow\infty}J=0\nn
\eeq

where $M_1,M_2$ are the initial and final Cauchy surfaces that asymptote to constant time slices.

If the third term vanishes then the bulk symplectic structure is already hypersurface independent. If the third term does not vanish and turns out to be exact i.e,

\beq
\int_{r\rightarrow\infty}J=\int_{r\rightarrow\infty}dj
\eeq
then the symplectic structure given by $\int_MJ-\int_{S_\infty}j$ (where $S_\infty$ is the two surface at the intersection of the hypersurface  $M$ with the boundary) is hypersurface independent and $j(\delta_1,\delta_2)$ is the ``Boundary symplectic current''.  The hypersurface independent symplectic structure is then given by:
\beq
\tilde{\Omega}(\delta_1,\delta_2)&=&-\frac{1}{8\pi 
G}\int_{M}\left(\delta_{[1}\Sigma_{IJ}\wedge\delta_{2]} \omega^{IJ}\right)-\int_{S_\infty}j(\delta_1,\delta_2)\nn
\eeq

For a vector field $X$ the variation $\delta_X$ acts on the fields like a lie derivative $\lie_X$. One can then show that if the equations of motion hold in the bulk, then the bulk symplectic structure $\Omega(\delta,\delta_X)$ contributes only at the boundary $\partial M$ of the cauchy surface $M$. Therefore it follows that,

\beq
\tilde{\Omega}(\delta,\delta_{X})&=&\Omega(\delta,\delta_{X})-\int_{S_\infty}j(\delta,\delta_X)
\eeq
where
\beq
\Omega(\delta,\delta_{X})=-\frac{1}{16\pi G}\int_{\partial M}[(X.\omega^{IJ})\delta\Sigma_{IJ}-
(X.\Sigma_{IJ})\wedge \delta \omega^{IJ}]\nn
\eeq

For another vector field $X'$ it immediately follows that,

\beq\label{eq:alg}
\tilde{\Omega}(\delta_{X'},\delta_{X})&=&\Omega(\delta_{X'},\delta_{X})-\int_{S_\infty}j(\delta_{X'},\delta_X)
\eeq
where
\beq
\Omega(\delta_{X'},\delta_{X})&=&\nn
-\frac{1}{16\pi G}\int_{\partial M}&&[(X.\omega^{IJ})\lie_{X'}\Sigma_{IJ}-
(X.\Sigma_{IJ})\wedge \lie_{X'}\omega^{IJ}]\nn
\eeq
It then implies that if the vector fields are 
Hamilonian(sec. \ref{INTCHARGES}).
\beq
[H_X,H_{X'}]&=&H_{[X,X']}+\tilde{\Omega}(\delta_{X'},\delta_{X})
\eeq
where the term $H_{[X,X']}$ is added to take into account the non vanishing of $[\delta_X,\delta_{X'}]$.

%%%%%%%%%%%%%%%%%%%%%%%%%%%%%%%%%%%%%%%%%%%%%%%%%%%%%%%%%%%%%%%%%%%%%%%%%%%%%%%%%%%%%%%%%%%%%%%%%%%%%%%%%%%%%%%%%%%%%%%%%%%%%%%%%%%%%%%%%%%%%%%%%%%%%%%%%%%%%%%%%%%%%%%%%%%%%%%%%%%%%%%%%%%%%%%%%%%%%%%%%%%%%%%%%%%%%%%%%%%%%%%%%%%%%%%%%%%%%%%%%%%%%%%%%%%%%%%%%%%%%%%%%%%%%%%%%%%%%%%%%%%%%%%%%%%%%%%%%%%%%%%%%%%%
\subsection{The Boundary Symplectic structure}\label{app:SSS}

To go ahead with any calculation we first need to find the boundary symplectic structure. The only contributions to the boundary symplectic structure come from $\delta_{[1}\Sigma_{10}\wedge\delta_{2]}\omega^{10}$ and $\delta_{[1}\Sigma_{30}\wedge\delta_{2]}\omega^{30}$. For variations of the form (which corresponds to variations about the ANHEK background obeying the linearized asymptotic constraints).

\beq\label{constraint}
A&=&C=1\nn
B&=&0\nn
\delta A&=&\delta C\nn
\delta B&=&\partial_\phi{\delta A}
\eeq

and using the form of the connection calculated from only the zeroth order tetrad (Appendix \ref{conn}) one can show that,
\beq
\int_{r\rightarrow\infty}J&=&\frac{1}{4\pi G}\int \frac{\partial}{\partial t}\left(\frac{N^2\Lambda}{A}\delta_{[1}A\delta_{2]}B\right) dt\wedge d\theta\wedge d\phi\nn
&=&\frac{1}{4\pi G}\int_{S_2}\left(\frac{N^2\Lambda}{A}\delta_{[1}A\delta_{2]}B\right)d\theta\wedge d\phi\nn
&&~~~~~~~-\frac{1}{4\pi G}\int_{S_1}\left(\frac{N^2\Lambda}{A}\delta_{[1}A\delta_{2]}B\right)d\theta\wedge d\phi\nn
\eeq
where $S_1,S_2$ are the intersections of $M_1,M_2$ respectively with the boundary.

To arrive at the above result we first identified the total time derivative and then used restrictions eq. (\ref{constraint}) to see if the other terms vanish.

It then follows that the relevant hypersurface independent quantity $\tilde{\Omega}(\delta,\delta_X)$ is given by
\beq
\tilde{\Omega}(\delta,\delta_X)&=&\nn
&&-\frac{1}{16\pi G}\int_{S_\infty}~~[(X.\omega^{IJ})\delta\Sigma_{IJ}-
(X.\Sigma_{IJ})\wedge \delta 
\omega^{IJ}]\nn
&&-\frac{1}{8\pi G}\int_{S_\infty}\frac{N^2\Lambda}{A}\left(\delta A~\delta_X B-\delta_X A~\delta B\right)d\theta\wedge d\phi\nn
\eeq

In general the boundary symplectic structure can have non-zero order one contributions coming from higher order terms in the asymptotic expansion. Ideally one should do the asymptotic expansion and check the boundary symplectic structure order by order. We must point out that we were unable to find a systematic way to isolate the terms of different orders in $\omega^{IJ}$. However in this case since we will be studying perturbations generated by $\xi$ around the ANHEK background it would suffice to check the zeroth order tetrad.

%%%%%%%%%%%%%%%%%%%%%%%%%%%%%%%%%%%%%%%%%%%%%%%%%%%%%%%%%%%%%%%%%%%%%%%%
%%%%%%%%%%%%%%%%%%%%%%%%%%%%%%%%%%%%%%%%%%%%%%%%%%%%%%%%%%%%%%%%%%%%%%%%%
%%%%%%%%%%%%%%%%%%%%%%%%%%%%%%%%%%%%%%%%%%%%%%%%%%%%%%%%%%%%%%%%%%%%%%%%
%%%%%%%%%%%%%%%%%%%%%%%%%%%%%%%%%%%%%%%%%%%%%%%%%%%%%%%%%%%%%%%%%%%%%%%
%%%%%%%%%%%%%%%%%%%%%%%%%%%%%%%%%%%%%%%%%%%%%%%%%%%%%%%%%%%%%%%%%%%%%%%
%%%%%%%%%%%%%%%%%%%%%%%%%%%%%%%%%%%%%%%%%%%%%%%%%%%%%%%%%%%%%%%%%%%%%%%
%%%%%%%%%%%%%%%%%%%%%%%%%%%%%%%%%%%%%%%%%%%%%%%%%%%%%%%%%%%%%%%%%%%%%%%%

%%%%%%%%%%%%%%%%%%%%%%%%%%%%%%%%%%%%%%%%%%%%%%%%%%%%%%%%%%%%%%%%%%%%%%%%%%%%%%%%
%%%%%%%%%%%%%%%%%%%%%%%%%%%%%%%%%%%%%%%%%%%%%%%%%%%%%%%%%%%%%%%%%%%%%%%%%%%%%%%%

\subsection{Algebra of Charges}

We therefore go ahead and calculate $\tilde{\Omega}(\delta_\xi,\delta_{\xi'})$. To calculate the contribution from the bulk it is enough to consider only the NHEK tetrad (eq. \ref{backT}) and connection (eq. \ref{backConn}) and not the quantities in the asymptotic expansion. As can be seen that the relevant vector field has a non zero interior product $\xi.e^I$ for $I=1 $ and $3$.

\beq
\xi.e^1=-\frac{Nr\epsilon'}{\sqrt{1+r^2}}~~~~~~~~~~~~~~~~~~~~
\xi.e^3=\Lambda N\epsilon
\eeq

We note that $\xi.(\mathcal{X}\wedge \mathcal{Y})=(\xi.\mathcal{X})\mathcal{Y}-\mathcal{X}(\xi.\mathcal{Y})$ for one forms $\mathcal{X}$ and $\mathcal{Y}$. It then follows that,
 $\xi.\Sigma_{IJ}$ restricted to the two surfaces spanned by $\theta$ and $\phi$ survive only for,
\beq
\xi.\Sigma_{10}=\Lambda N^2\epsilon~ d\theta~~~~~~~~~~&&
\xi.\Sigma_{20}=\Lambda^2 N^2\epsilon~d\phi \nn
\nn
\xi.\Sigma_{30}=-\frac{N^2r\epsilon'}{\sqrt{1+r^2}}~d\theta~~~~&&
\eeq

The non zero terms for ~$\xi.\omega^{IJ}$ can be readily calculated from the expression of the connection.
\beq
\xi.\omega^{10}&=&\frac{1}{2}\Lambda^2\epsilon\nn
\nn
\xi.\omega^{30}&=&-\frac{1}{2}\frac{\Lambda}{\sqrt{1+r^2}}r\epsilon'\nn
\nn
\xi.\omega^{32}&=&\frac{2(\Lambda N)'}{N}\epsilon
\eeq

To calculate~$\lie_{\xi}\Sigma_{IJ}$ one uses the expression for the action of lie derivative on forms
\beq
\lie_{\xi}\Sigma_{IJ}&=&d(\xi.\Sigma_{IJ})+\xi.d\Sigma_{IJ}
\eeq
On restricting the two form to the two surface spanned by $\theta$ and $\phi$
one gets the following non zero components.
\beq
\lie_{\xi}\Sigma_{10}&=&\Lambda N^2 
\epsilon'd\phi\wedge d\theta\nn
\nn
\lie_{\xi}\Sigma_{30}&=&-\frac{N^2r\epsilon''}{\sqrt{1+r^2}}d\phi\wedge d\theta\nn
\eeq

$\lie_{\xi}\omega^{IJ}$ can be similarly be calculated and their restriction  to the two surfaces have the following form,
\beq
\lie_{\xi}\omega^{10}&=&\frac{1}{2}\Lambda^2\epsilon'd\phi\nn
\nn
\lie_{\xi}\omega^{30}&=&-\frac{1}{2}\frac{\Lambda r\epsilon''}{\sqrt{1+r^2}}d\phi
\eeq

Having calculated all the required terms one can go ahead and calculate $\Omega(\delta_{\xi},\delta_{\xi'})$. Putting everything together one gets,

\beq
\Omega(\delta_{\xi_m},\delta_{\xi_n})=&\nn
\frac{1}{8\pi G}\int_{S_{\infty}}&\left[\Lambda^3 N^2\epsilon_m\epsilon_n'd\theta\wedge
d\phi+N^2\Lambda\epsilon_m'\epsilon_n''d\theta\wedge d\phi\right]\nn
\eeq

with the substitution $\epsilon_m=-e^{-im\phi}$ as in \cite{Guica:2008mu} we get,\\
\beq
\nn
\Omega(\delta_{\xi_m},\delta_{\xi_n})=&&\nn
\nn
\frac{i(m)\delta_{m+n,0}}{4G}\int_{S_{
\infty}}&&\Lambda^3
N^2d\theta+\frac{i(mn^2)\delta_{m+n,0}}{4G}\int_{S_{\infty}}N^2\Lambda d\theta\nn
\eeq
The relevant integrals can be calculated and are given as:
\beq
\int\Lambda^3N^2d\theta&=&2JG\int_{0}^{\pi}\frac{4sin^2\theta}{(1+cos^2\theta)^2
}d\theta=8JG\nn
\nn
\int\Lambda N^2d\theta&=&2JG\int_{0}^{\pi}sin\theta d\theta=4JG
\eeq
Therefore,\\
\beq
\Omega(\delta_{\xi_m},\delta_{\xi_n})&=&i(m^3+2m)J\delta_{m+n,0}
\eeq

We also need to check whether $j(\delta_\xi,\delta_{\xi'})$ contributes to the central charge. Using the variations Appendix \ref{variation} we see that for the ANHEK background,
\beq
\int_{S_\infty}j(\delta_{\xi_m},\delta_{\xi_n})&=&\frac{1}{8\pi G}\int_{S_\infty}N^2\Lambda(\epsilon_m'\epsilon_n''-\epsilon_n'\epsilon_m'')d\theta\wedge d\phi\nn
&=&2Jm^3\delta_{m+n,0}
\eeq
Therefore it follows that,
\beq
\tilde{\Omega}(\delta_{\xi_m},\delta_{\xi_n})&=&i(-m^3+2m)J\delta_{m+n,0}
\eeq
%%%%%%%%%%%%%%%%%%%%%%%%%%%%%%%%%%%%%%%%%%%%%%%%%%%%%%%%%%%%%%%%%%%%%%%%%%%%%%%%%%%%%%%%%%%%%%%%%%%%%%%%%%%%%%%%%%%%%%%%%%%%%%%%%%%%%%%%%%%%%%%%%%%%%%%%%%%%%%%%%%%%%%%%%%%%%%%%%%%%%%%%%%%%%%%%%%%%%%%%%%%%%%%%%%%%%%%%%%%%%%%%%%%%%%%%%%%%%%%%%%%%%%%%%%%%%%%%%%%%%%%%%%%%%%%%%%%%%%%%%%%%%%%%%%%%%%%%%%%%%%%%%%%%%%%%%%%%%%%%%%%%%%%%%%%%%%%%%%%%%%%%%%%%%%%%%%%%%%%%%%%%%%%%%%%%%%%%%%%%

\subsection{Hamiltonian}\label{INTCHARGES}
To see if the vector fields are Hamiltonian we check, 
whether $\tilde{\Omega}(\delta,\delta_\xi)$ can be written as a total variation. We do this in two steps. First we consider only the bulk symplectic structure $\Omega(\delta_1,\delta_2)$ and then check the contributions from $j(\delta_1,\delta_2)$.

We note that here we need the asymptotic expansions. The only terms that will contribute to the expression of $\Omega({\delta,\delta_\xi})$ are then seen to be I=1, J=0 and I=3, J=0. The relevant terms restricted to the two surface is then of the form:
\beq
\omega^{10}&=&g_1(t,\theta,\phi)d\phi+g_2(t,\theta,\phi)d\theta\nn
\nn
\omega^{30}&=&h_1(t,\theta,\phi)d\phi+h_2(t,\theta,\phi)d\theta\nn
\nn
\xi.\omega^{10}&=&g_1(t,\theta,\phi)\epsilon(\phi)\nn
\nn
\xi.\omega^{30}&=&-\frac{1}{2}(\Lambda)\epsilon'(\phi)+h_1(t,\theta,\phi)\epsilon(\phi)
\eeq
where $g_{1,2}(t,\theta,\phi),h_{1,2}(t,\theta,\phi)$ are functions which depend on $\Lambda,\Omega,A,B,C$ and their derivatives. First we consider the bulk symplectic structure,
\beq
\Omega(\delta,\delta_{\xi})=-\frac{1}{16\pi G}\int_{\partial 
M}~~[(\xi.\omega^{IJ})\delta\Sigma_{IJ}-\nn
(\xi.\Sigma_{IJ})\wedge \delta 
\omega^{IJ}]
\eeq
We note that,
\beq
(\xi.\omega^{10})\delta\Sigma_{10}-(\xi.\Sigma_{10})\delta\omega^{10}&=&\nn
\nn
g_1\epsilon\delta\left(\frac{N^2\Lambda}{A}\right) d\theta\wedge d\phi
&+&\left(\frac{N^2\Lambda\epsilon}{A}\right)\delta g_1 ~d\theta\wedge d\phi\nn
\eeq
\beq
(\xi.\omega^{30})\delta\Sigma_{30}-(\xi.\Sigma_{30})\delta\omega^{30}=&\nn
\nn
(-\frac{1}{2}\Lambda\epsilon'+h_1\epsilon)\delta(N^2B)d\phi\wedge d\theta
\nn
-(-N^2\epsilon'+&N^2B\epsilon)\delta h_1~d\theta\wedge d\phi\nn
\eeq
It is therefore at once evident  that the contribution from the bulk symplectic structure is integrable provided we assume $\delta\epsilon=0$.

For the vector fields $\xi$ in question, we also need to check if the charges are still integrable with the addition of the boundary symplectic current. Using the expressions for $\delta_\xi A, \delta_\xi B, \delta_\xi C$ from Appendix \ref{variation} we see that this contribution is equal to,
\beq
\frac{1}{A}&&\left(\delta_\xi A\delta B-\delta_\xi B\delta A\right)\nn
&&=\frac{1}{A}\left(-\epsilon'A+\epsilon\partial_\phi A\right)\delta B-\frac{1}{A}\left(-\epsilon''+\epsilon\partial_\phi B+\epsilon'B\right)\delta A\nn
\eeq
We note that the first and the third term are integrable. So we concentrate on the other terms.
\beq
\epsilon\partial_\phi&&\left(\log{A}\right)\delta B-\epsilon\partial_\phi B\delta\left(\log{A}\right)-\epsilon'B\delta\left(\log{A}\right)\nn
\nn
&&=\partial_\phi\left[\epsilon\left(\log{A}\right)\delta B\right]-\epsilon\left(\log{A}\right)\delta \partial_\phi B\nn
\nn
&&~~~~-\epsilon'\left(\log{A}\right)\delta B-\epsilon\partial_\phi B\delta\left(\log{A}\right)-\epsilon'B\delta\left(\log{A}\right)\nn
\nn
&&\equiv-\delta\left(\epsilon\left(\log{A}\right)\partial_\phi B\right)-\delta\left(\epsilon'B\left(\log{A}\right)\right)
\eeq

where we have omitted the first term, while going from first to second expression, as it is a total $\phi$ derivative and does not contribute to the integral. So it follows that the charges are still integrable.  Moreover for the ANHEK background (for which $A=1$ and $B=0$) the boundary symplectic structure  does not contribute to the Hamiltonian. So, for the given background one can set the Hamiltonian function to be,
\beq
H_\xi=\left[-\frac{1}{16\pi G}\int_{\partial 
M}~~N^2\Lambda^3\epsilon d\theta\wedge d\phi\right]
\eeq

%%%%%%%%%%%%%%%%%%%%%%%%%%%%%%%%%%%%%%%%%%%%%%%%%%%%%%%%%%%%%%%%%%%%%%%%%%%%%%%%
%%%%%%%%%%%%%%%%%%%%%%%%%%%%%%%%%%%%%%%%%%%%%%%%%%%%%%%%%%%%%%%%%%%%%%%%%%%%%%%%
%%%%%%%%%%%%%%%%%%%%%%%%%%%%%%%%%%%%%%%%%%%%%%%%%%%%%%%%%%%%%%

\subsection{Entropy calculations}\label{entropy}
To calculate the entropy we choose an approach outlined in \cite{silva} and used in \cite{Dreyer:2013noa}. The charge for the vector field $\xi$ has been calculated in sec \ref{INTCHARGES}. It therefore follows that
\beq
H_{[\xi_m,\xi_n]}&=&-\frac{1}{16\pi G}\int_{\partial 
M}~~N^2\Lambda^3(\epsilon_m\epsilon_n'-\epsilon_n\epsilon_m') d\theta\wedge d\phi\nn
\nn
&=&-\frac{1}{16\pi G}i(m-n)2\pi\delta_{m+n,0}\times 8\pi G\nn
\nn
&=&-2imJ\delta_{m+n,0}\nn
\eeq
putting this in the expression for Poisson Bracket, we get:
\beq
[H_{\xi_m},H_{\xi_n}]&=&-iJm^3\delta_{m+n,0}\nn
\nn
i[H_{\xi_m},H_{\xi_n}]&=&Jm^3\delta_{m+n,0}
\eeq

Comparing this with the virasoro algebra.
\beq
i[H_{m},H_{n}]&=&(m-n)H_{m+n}+\frac{c}{12}(m^3-m)\delta_{m+n,0}\nn
\eeq
It then follows that,
\beq
i[H_{1},H_{-1}]&=&2H_0=J\nn
\nn
i[H_{2},H_{-2}]&=&4H_0+\frac{c}{2}=8J
\eeq

One can now solve the above system of linear algebraic equations for $c$ and $H_0$, which gives:
\beq
c=12J~~~~~&&~~~~~H_0=\frac{J}{2}
\eeq
Now using Cardy formula:
\beq
S=2\pi\sqrt{\frac{cH_0}{6}}=2\pi J
\eeq
which is in accordance with the Bekenstein-Hawking Entropy formula. The Planck's constant in the formula can be recovered by the naive quantization $i\hslash[H_{m},H_{n}]\rightarrow [H_{m},H_{n}]$.

%%%%%%%%%%%%%%%%%%%%%%%%%%%%%%%%%%%%%%%%%%%%%%%%%%%%%%%%%%%%%%%%%%%%%%%%%%%%%%%%
%%%%%%%%%%%%%%%%%%%%%%%%%%%%%%%%%%%%%%%%%%%%%%%%%%%%%%%%%%%%%%%%%%%%%%%%%%%%%%%%
%%%%%%%%%%%%%%%%%%%%%%%%%%%%%%%%%%%%%%%%%%%%%%%%%%%%%%%%%%%%%%%%%%%%%%%%%%%%%%%%
%%%%%%%%%%%%%%%%%%%%%%%%%%%%%%%%%%%%%%%%%%%%%%%%%%%%%%%%%%%%%%%%%%%

\section{Holst Action}
In the first order formulation, both the Holst and Palatini actions give the same equations of motion, viz. Einstein's equations in spite of the fact that the two actions differ by a term which is not a total derivative. Therefore, NHEK is a solution of both these actions. It is therefore legitimate to check whether under NHEK boundary conditions the use of Holst action gives a different result from the Palatini action.

The Holst action in the bulk is given by:

\beq 
S_{H} = - \frac{1}{16\pi G} \int_\mathcal{M} \Sigma_{IJ}
\wedge \left(F^{IJ} + \frac{1}{\gamma}~ ^*F^{IJ}\right)
\eeq
where $^*F^{IJ}=\frac{1}{2}\epsilon^{IJ}~_{KL}F^{KL}$, $\gamma$ is the Immirzi parameter.

The symplectic current is then given by:
\\
\beq
\label{J} J_H(\delta_1, \delta_2) = -\frac{1}{8\pi G} \left[ \delta_{[1} \Sigma_{IJ} \wedge
\delta_{2]} \left( \omega^{IJ} + \frac{1}{\gamma}~^* \omega^{IJ} \right)\right]\nn
\eeq

On half-shell i.e if the torsion free conditions holds then the symplectic current simplifies \cite{Corichi:2010ur} and is then given by:
\beq
J_{H}(\delta_1,\delta_2)=&&\nn
-\frac{1}{16\pi G}&&
(\delta_{[1}\Sigma_{IJ}\wedge\delta_{2]} \omega^{IJ})+\frac{1}{8\pi G\gamma}d(\delta_{[1}e_I\wedge\delta_{2]}e^I)\nn
\eeq

We first note that the first term in the above expression is the usual Palatini term (denoted by $J_p$ in the next expression). To construct the hypersurface independent symplectic structure we note that on shell:

\beq
d J_{H}=0
\eeq
this implies that when integrated over a closed region of spacetime bounded by $M_1\cup M_2\cup B$ (where B is a portion of the boundary of spacetime given by $r\rightarrow \infty$ in our case):
\beq
\left(\int_{M_1}-\int_{M_2}+\int_{r\rightarrow\infty}\right)J_{P}&&\nn
+\left(\int_{M_1}-\int_{M_2}+\int_{r\rightarrow\infty}\right)&&d(\delta_{[1}e_I\wedge\delta_{2]}e^I)=0\nn
\eeq
where $M_1,M_2$ are the initial and final Cauchy surfaces that asymptote to constant time slices.

We note that the second term is always zero. So the Immirzi parameter can never appear in the hypersurface independent symplectic structure calculated from Holst action. So the Holst term modifies neither the poisson bracket nor the Hamiltonian no matter what geometry or boundary conditions one is considering. So if one uses the Holst action instead of the Palatini action the sem-classical entropy is still the same as that calculated from Palatini action and is therefore independent of the Immirzi parameter.

%\subsection{Algebra of charges}
%We note that the first term in eq.(\ref{eq:HC}) is the usual Palatini term which has already been calculated in the previous sections. We therefore need to check if the second term contributes to the algebra of charges for the vector fields $\xi$. For the NHEK background the only terms that will contribute to this extra term are $I=1,3$.
%\beq
%\lie_\xi e^1=d(-N\epsilon')+N'\epsilon'd\theta=-N\epsilon''d\phi\nn
%\nn
%\lie_\xi e^3=d(N\Lambda\epsilon)+N'\Lambda'\epsilon d\theta=-N\Lambda\epsilon'd\phi
%\eeq 
%So it is evident that the Holst term  doesn't contribute.\\

%\subsection{Hamiltonian}
%The first term in eq.(\ref{eq:HH}) is the usual Palatini term. We therefore need to check if the second term contributes to the Hamiltonian for the vector fields $\xi$.
%The extra term coming from the Holst terms is:
%\beq
%\frac{1}{16\pi G\gamma}\int_{S_\infty}(\delta e_I\wedge\delta_{\xi}e^I-\delta_{\xi} e_I\wedge\delta e^I)
%\eeq

%Therefore there is no contribution to the Hamiltonian as well. So it is evident that the Holst term has no role to play in the symmetry based approach, as far as Kerr/CFT is concerned. So if one works with the Holst action instead of the Palatini action the expression for entropy does not change and therefore the Immirzi parameter does not enter the expression of entropy. 

\section{Discussion}

Apart from the motivations pointed out in the introduction, the first order formalism gives a cleaner calculation. For example, it is evident that the desired central extension comes from terms like $(\xi.\omega^{IJ})d(\xi.\Sigma_{IJ})$ and $d(\xi.\omega^{IJ})\wedge \xi.\Sigma_{IJ}$.
Therefore, from the expressions of the tetrads and connections it is possible to predict which vector field will give an $m^3$ term.

It seems that in the second order formulation the boundary symplectic structure has been studied only in the context of asymptotically flat geometries. Such studies have not been made in symmetry based approaches in the second order formulation. Therefore, the symplectic structure given in \cite{Barnich:2001jy} may not be hypersurface independent for the boundary conditions appropriate for the NHEK geometry. This has been pointed out for Kerr/CFT in \cite{Amsel:2009pu}.

In this case the boundary symplectic structure does not vanish. A non-vanishing boundary symplectic structure implies that the bulk symplectic structure alone is not hypersurface independent. This would precisely give a Hamiltonian calculated from the bulk symplectic structure to be hypersurface dependent. Since for the NHEK background and the vector fields generating the $Diff~S^1$, the Hamiltonian calculated from the bulk symplectic structure is already time independent, it was expected that atleast for $\xi$ the boundary symplectic structure should not contribute. However the results of section \ref{app:SSS} show that there is a non trivial contribution to the central charge from the boundary symplectic structure.

We show, by explicit calculation, that only if the boundary symplectic structure is taken into account i.e one works with a truly hypersurface independent symplectic  structure, the entropy results match with those obtained in second order formulation. So even though the results don't change we think that the relevance and importance of the boundary symplectic structure has been fully conveyed in this work.

\section{Conclusion}

We studied the Kerr-CFT correspondence using the symplectic structure in the first order formulation of gravity. The Boundary symplectic structure has been studied. It is shown that it  does not vanish. The results obtained are then in agreement with those already obtained in the second order formulation. We studied the effect of adding the Holst term and showed that it does not contribute.

It is known that the Immirzi parameter labels the nonequivalent quantisation in LQG. It is also believed that a fine tuning of the Immirzi parameter is required in order to reproduce the BH entropy formula. However, recently in \cite{Ghosh:2011fc} it has been argued that the Immirzi parameter is not so relevant in getting the semiclassical value for BH entropy. 
Our result that the entropy formula is independent of the Immirzi parameter is consistent with the claim that it plays no fundamental role in the quantum theory.

It will be interesting though to see if the Immirzi parameter contributes to Wald entropy for NHEK. Wald prescription for the black hole entropy works only for bifurcate Killing horizons. Hence, a straightforward implementation of this method to the case of extremal Kerr is not possible. A widely accepted approach is to calculate the entropy for a non-extremal black hole and then take the extremal limit or along the lines of \cite{Hajian:2013lna} for instance.

\section{acknowledgements}
The author wants to thank Amit Ghosh for useful discussions, David Sloan for clarifying a point from a reference and Alejandro Corichi for a useful clarification.

\appendix
%%%%%%%%%%%%%%%%%%%%%%%%%%%%%%%%%%%%%%%%%%%%%%%%%%%%%%%%%%%%%%%%%%%%%%%%%%%%%%%%
%%%%%%%%%%%%%%%%%%%%%%%%%%%%%%%%%%%%%%%%%%%%%%%%%%%%%%%%%%%%%%%%%%%%%%%%%%%%%%%%
%%%%%%%%%%%%%%%%%%%%%%%%%%%%%%%%%%%%%%%%%%%%%%%%%%%%%%%%%%%%%%%%%%%%%%%%%%%%%%%%
%%%%%%%%%%%%%%%%%%%%%%%%%%%%%%%%%%%%%%%%%%%%%%%%%%%%%%%%%%%%%%%%%%%%%%%%%%%%%%%%
%%%%%%%%%%%%%%%%%%%%
%%%%%%%%%%%%%%%%%%%%%%%%%%%%%%%%%%%%%%%%%%%%%%%%%%%%%%%%%%%%%%%%%%%%%%%%%%%%%%%%%%%%%%%%%%%%%%%%%%%%%%%%%%%%%%%%%%%%%%%%%%%%%%%%%%%%%%%%%%%%%%%%%%%%%%%%%%%%%%%%%%%%%%%%%%%%%%%%%%%%%%%%%%%%%%%%%%%%%%%%%%%%%%%%%%%%%%%%%%%%%%%%%%%%%%%%%%%%%%%%%%%%%%%%%%%%%%%%%%%%%%%%%%%%%%%%%%%%%%%%%%%%%%%%%%%%%%%%%%%%%%%%

%\section{The vector field $\xi$}\label{vect}
%In the Poincare type coordinate system $(y,\tau,\varphi)$ given by:
%\beq
%y&=&\left(\cos t\sqrt{1+r^2}+r\right)^{-1}\nn
%\tau&=&y\sin t\sqrt{1+r^2}\nn
%\varphi&=&\phi+\log\frac{\cos t+r\sin t}{1+\sin t\sqrt{1+r^2}}
%\eeq
%In the limit $r\rightarrow\infty$
%\beq
%\bar{\xi}^y&=&\xi^r\frac{\partial y}{\partial r}+\xi^t\frac{\partial y}{\partial t}+\xi^\phi\frac{\partial y}{\partial \phi}\nn
%&=&\frac{r\epsilon'}{r^2(1+\cos t)}\rightarrow 0.
%\eeq
%So the vector field $\xi$ is indeed tangent to the boundary $y\rightarrow 0$ or  $r\rightarrow\infty$.
%%%%%%%%%%%%%%%%%%%%%%%%%%%%%%%%%%%%%%%%%%%%%%%%%%%%%%%%%%%%%%%%%%%%%%%%%%%%%%%%
%%%%%%%%%%%%%%%%%%%%%%%%%%%%%%%%%%%%%%%%%%%%%%%%%%%%%%%%%%%%%%%%%%%%%%%%%%%%%%%%
%%%%%%%%%%%%%%%%%%%%%%%%%%%%%%%%%%%%%%%%%%%%%%%%%%%%%%%%%%%%%%%%%%%%%%%%%%%%%%%%
%%%%%%%%%%%%%%%%%%%%%%%%%%%%%%%%%%%%%%%%%%%%%%%%%%%%%%%%%%%%%%%%%%%%%%%%%%%%%%%%
%%%%%%%%%%%%%%%%%%%%
\section{Variations of $A,B$ an $C$}\label{variation}
Note that $\delta_\xi A$ is not equal to $\lie_\xi A$. Rather it has to be calculated from the action of $\lie_\xi$ on the fields.
\beq
\lie_\xi e^0=N(-r\epsilon'Adt+r\epsilon\partial_\phi A dt)
\eeq
It therefore follows that $\delta_\xi A=-\epsilon'A+\epsilon\partial_\phi A $.

Similarly
\beq
\lie_\xi e^3=N\Lambda\left(\frac{\epsilon'}{C}-\frac{\epsilon\partial_\phi C}{C^2}\right)d\phi+N\Lambda r(-\epsilon'C+\epsilon\partial_\phi C)dt\nn
\eeq 
which implies
\beq
\delta_\xi(\frac{1}{C})=\frac{\epsilon'}{C}-\frac{\epsilon\partial_\phi C}{C^2}
\eeq
Therefore $\delta_\xi C=-\epsilon'C+\epsilon\partial_\phi C$. For consistency one can check that the $dt$ term gives the same variation.

\beq
\lie_\xi e^1=N(B\epsilon'+\epsilon \partial_\phi B-\epsilon'')d\phi
\eeq
Therefore $\delta_\xi B=B\epsilon'+\epsilon \partial_\phi B-\epsilon''$

%%%%%%%%%%%%%%%%%%%%%%%%%%%%%%%%%%%%%%%%%%%%%%%%%%%%%%%%%%%%%%%%%%%%%%%%%%%%%%%%%%%%%%%%%%%%%%%%%%%%%%%%%%%%%%%%%%%%%%%%%%%%%%%%%%%%%%%%%%%%%%%%%%%%%%%%%%%%%%%%%%%%%%%%%%%%%%%%%%%%%%%%%%%%%%%%%%%%%%%%%%%%%%%%%%%%%%%%%%%%%%%%%%%%%%%%%%%%%%%%%%%%%%%%%%%%%%%%%%%%%%%%%%%%%%%%%%%%%%%%%%%%%%%%%%%%%%%%%%%%%%%%

\begin{widetext}
\section{Form of the connection}\label{conn}
The form of the connection calculated from the zeroth order tetrad is of the form,
\beq
~^0\omega^{10}=&& -\frac{1}{2}\,{\frac {2\, A \left( t,\theta,\phi \right)^{2}-\Lambda \left( \theta \right)^{2} C \left( t,\theta,\phi \right)^{2}}{A \left( t,\theta,\phi \right) }}rdt\nn
&&-\frac{1}{2}\,{\frac {C \left( t,\theta,\phi \right)^{2}{\frac {\partial }{\partial \theta}}B \left( t,\theta,\phi \right) }{A \left( t,\theta,\phi \right) }}d\theta+\frac{1}{2}\,{\frac {\Lambda \left( \theta \right) ^{2}}{A \left( t,\theta,\phi \right) }}d\phi\nn
\nn
~^0\omega^{20}=&&-{\frac {\left(  \left( {\frac {d}{d\theta}}N \left( \theta \right)  \right)  A \left( t,\theta,\phi \right)^{2}+N \left( \theta \right) A \left( t,\theta,\phi \right) {\frac {\partial }{\partial \theta}}A \left( t,\theta,\phi \right) -N \left( \theta \right)  \Lambda \left( \theta \right)^{2}C \left( t,\theta,\phi \right) {\frac {\partial }{\partial \theta}}C \left( t,\theta,\phi \right) \mbox{} \right) }{N \left( \theta \right) A \left( t,\theta,\phi \right) }}rdt\nn
&&-\frac{1}{2}\,{\frac { C \left( t,\theta,\phi \right)^{2}{\frac {\partial }{\partial \theta}}B \left( t,\theta,\phi \right)
\mbox{}}{A \left( t,\theta,\phi \right) }}\frac{dr}{r}+\frac{1}{2}\,{\frac {-B \left( t,\theta,\phi \right)  \left( {\frac {\partial }{\partial \theta}}B \left( t,\theta,\phi \right)  \right) C \left( t,\theta,\phi \right)^{3}
\mbox{}+2\, \Lambda \left( \theta \right)^{2}{\frac {\partial }{\partial \theta}}C \left( t,\theta,\phi \right) }{A \left( t,\theta,\phi \right) C \left( t,\theta,\phi \right) }}d\phi
\nn
\nn
~^0\omega^{30}=&&\frac{1}{{\Lambda \left( \theta \right) A \left( t,\theta,\phi \right) }} \left(B \left( t,\theta,\phi \right) C \left( t,\theta,\phi \right)A \left( t,\theta,\phi \right)^{2}-B \left( t,\theta,\phi \right)C \left( t,\theta,\phi \right)^{3} \Lambda \left( \theta \right)^{2}\right.\nn
&&~~~~~~~~~~~~~~~~~~~~~\left.\mbox{}-C \left( t,\theta,\phi \right) A \left( t,\theta,\phi \right) {\frac {\partial }{\partial \phi}}A \left( t,\theta,\phi \right)+C \left( t,\theta,\phi \right)^{2} \Lambda \left( \theta \right)^{2}{\frac {\partial }{\partial \phi}}C \left( t,\theta,\phi \right)\right)rdt\nn
\nn
&&~~~~~~~~~~~~~~~~~~~~~~~~~~~~~~+\frac{1}{2}\,{\frac {C \left( t,\theta,\phi \right)\Lambda \left( \theta \right)^{2}
\mbox{}}{\Lambda \left( \theta \right) A \left( t,\theta,\phi \right) }}\frac{dr}{r}+{\frac {\Lambda \left( \theta \right) {\frac {\partial }{\partial \theta}}C \left( t,\theta,\phi \right) }{A \left( t,\theta,\phi \right) }}d\theta\nn
\nn
&&-\frac{1}{2}\,\frac{1}{C \left( t,\theta,\phi \right)^{2}\Lambda \left( \theta \right) A \left( t,\theta,\phi \right) } \left(-2\,C \left( t,\theta,\phi \right)^{2} \Lambda \left( \theta \right)^{2}{\frac {\partial }{\partial \phi}}C \left( t,\theta,\phi \right) 
\mbox{}+B \left( t,\theta,\phi \right) C \left( t,\theta,\phi \right)^{3} \Lambda \left( \theta \right)^{2}
\mbox{}\right) d\phi\nn
\nn
\nn
~^0\omega^{21}=&&-{\frac {{\frac {d}{d\theta}}N \left( \theta \right) }{N \left( \theta \right) r}}dr-\frac{1}{2}\,{\frac {2\,B \left( t,\theta,\phi \right) {\frac {d}{d\theta}}N \left( \theta \right) +N \left( \theta \right) {\frac {\partial }{\partial \theta}}B \left( t,\theta,\phi \right) }{N \left( \theta \right) }}d\phi
\nn
\nn
~^0\omega^{31}=&&\frac{1}{2}\,{\frac {C \left( t,\theta,\phi \right)\Lambda \left( \theta \right)^{2}}{\Lambda \left( \theta \right) }}rdt+\frac{1}{2}\,{\frac {C \left( t,\theta,\phi \right) {\frac {\partial }{\partial \theta}}B \left( t,\theta,\phi \right) }{\Lambda \left( \theta \right) }}d\theta\nn
\nn
\nn
~^0\omega^{32}=&&{\frac {C \left( t,\theta,\phi \right) \left( \Lambda \left( \theta \right) {\frac {d}{d\theta}}N \left( \theta \right) +N \left( \theta \right) {\frac {d}{d\theta}}\Lambda \left( \theta \right)  \right)
\mbox{}}{N \left( \theta \right) }}rdt+\frac{1}{2}\,{\frac {C \left( t,\theta,\phi \right) {\frac {\partial }{\partial \theta}}B \left( t,\theta,\phi \right) }{\Lambda \left( \theta \right) }}\frac{dr}{r}\nn
\nn
&&+\frac{1}{2}\,\frac{1}{ C \left( t,\theta,\phi \right)^{2}N \left( \theta \right) \Lambda \left( \theta \right) } \left(N \left( \theta \right) B \left( t,\theta,\phi \right)  \left( {\frac {\partial }{\partial \theta}}B \left( t,\theta,\phi \right)  \right) C \left( t,\theta,\phi \right)^{3}
\mbox{}+2\, \left( {\frac {d}{d\theta}}N \left( \theta \right)  \right) C \left( t,\theta,\phi \right)  \Lambda \left( \theta \right) ^{2}\right.\nn
&&~~~~~~~~~~~~~~~~~~~~~~~~~~~~~~~~~~~~\left.+2\,N \left( \theta \right) \Lambda \left( \theta \right)  \left( {\frac {d}{d\theta}}\Lambda \left( \theta \right)  \right) C \left( t,\theta,\phi \right) -2\,N \left( \theta \right) \Lambda \left( \theta \right)^{2}{\frac {\partial }{\partial \theta}}C \left( t,\theta,\phi \right) \right)d\phi
\eeq

\beq
~^1\omega^{10}&=&-\frac{1}{2}\frac{C \left( t,\theta,\phi \right)^{2}{\frac {\partial }{\partial t}}B \left( t,\theta,\phi \right) }{A \left( t,\theta,\phi \right)}rdt+\frac{1}{2}\,{\frac {-{\frac {\partial }{\partial t}}B \left( t,\theta,\phi \right)}{A \left( t,\theta,\phi \right) }}d\phi\nn
\nn
~^1\omega^{30}&=&\frac{\Lambda \left( \theta \right){\frac {\partial }{\partial t}}C \left( t,\theta,\phi \right)}{A \left( t,\theta,\phi \right) }rdt+\frac{1}{2}\,{\frac {C \left( t,\theta,\phi \right)  \left( -{\frac {\partial }{\partial t}}B \left( t,\theta,\phi \right)\right)
\mbox{}}{\Lambda \left( \theta \right)A \left( t,\theta,\phi \right) }}\frac{dr}{r}\nn
&&-\frac{1}{2}\,\frac{1}{ \left( C \left( t,\theta,\phi \right)  \right) ^{2}\Lambda \left( \theta \right) A \left( t,\theta,\phi \right) } \left(B \left( t,\theta,\phi \right)  \left( {\frac {\partial }{\partial t}}B \left( t,\theta,\phi \right)  \right)C \left( t,\theta,\phi \right)^{3}-2\,\Lambda \left( \theta \right)^{2}{\frac {\partial }{\partial t}}C \left( t,\theta,\phi \right)\right) d\phi\nn\nn
~^1\omega^{31}&=&\frac{1}{2}\,{\frac {C \left( t,\theta,\phi \right)  \left({\frac {\partial }{\partial t}}B \left( t,\theta,\phi \right)  \right) }{\Lambda \left( \theta \right) }}rdt
\eeq
\end{widetext}


\begin{thebibliography}{99}



\bibitem{Guica:2008mu} 
  M.~Guica, T.~Hartman, W.~Song and A.~Strominger,
  \emph{The Kerr/CFT Correspondence},
  Phys.\ Rev.\ D {\bf 80}, 124008 (2009)
  [arXiv:0809.4266 [hep-th]].
  
  
    \bibitem{Wu:2009di} 
  X.~-N.~Wu and Y.~Tian,
  \emph{Extremal Isolated Horizon/CFT Correspondence}
  Phys.\ Rev.\ D {\bf 80}, 024014 (2009)
  [arXiv:0904.1554 [hep-th]].
  
  C.~-Y.~Zhang, Y.~Tian and X.~-N.~Wu,
  \emph{Generalized Kerr/CFT correspondence with electromagnetic field},
  arXiv:1311.3384 [gr-qc].
  
  
  \bibitem{Amsel:2009pu} 
  A.~J.~Amsel, D.~Marolf and M.~M.~Roberts,
  \emph{On the Stress Tensor of Kerr/CFT}
  JHEP {\bf 0910}, 021 (2009)
  [arXiv:0907.5023 [hep-th]].
  
    \bibitem{Bardeen:1999px} 
  J.~M.~Bardeen and G.~T.~Horowitz,
  \emph{The Extreme Kerr throat geometry: A Vacuum analog of AdS(2) x S**2},
  Phys.\ Rev.\ D {\bf 60}, 104030 (1999)
  [hep-th/9905099].
  
  
  
  \bibitem{Brown:1986nw} 
  J.~D.~Brown and M.~Henneaux,
  \emph{Central Charges in the Canonical Realization of Asymptotic Symmetries: 
An Example from Three-Dimensional Gravity},
  Commun.\ Math.\ Phys.\  {\bf 104}, 207 (1986)
  
  
  \bibitem{symmetry}
A.~Strominger,
  \emph{Black hole entropy from near horizon microstates},
  JHEP {\bf 9802}, 009 (1998)
  [hep-th/9712251].
  
  
S.~Carlip,
  \emph{Black hole entropy from conformal field theory in any dimension},
  Phys.\ Rev.\ Lett.\  {\bf 82}, 2828 (1999)
  [hep-th/9812013].


S.~Carlip,
  \emph{Entropy from conformal field theory at Killing horizons},
  Class.\ Quant.\ Grav.\  {\bf 16}, 3327 (1999)
  [gr-qc/9906126].


S.~Carlip,
  \emph{Extremal and nonextremal Kerr/CFT correspondences},
  JHEP {\bf 1104}, 076 (2011)
  [Erratum-ibid.\  {\bf 1201}, 008 (2012)]
  [arXiv:1101.5136 [gr-qc]].
  
  
\bibitem{Barnich:2001jy} 
  G.~Barnich and F.~Brandt,
  \emph{Covariant theory of asymptotic symmetries, conservation laws and central charges}
  Nucl.\ Phys.\ B {\bf 633}, 3 (2002)
  [hep-th/0111246].
  
  
\bibitem{Dreyer:2013noa} 
  O.~Dreyer, A.~Ghosh and A.~Ghosh,
  \emph{Entropy from near-horizon geometries of Killing horizons}
  Phys.\ Rev.\ D {\bf 89}, 024035 (2014)
  [arXiv:1306.5063 [gr-qc]].
  




\bibitem{Ashtekar:2008jw}
   A.~Ashtekar, J.~Engle and D.~Sloan,
   \emph{Asymptotics and Hamiltonians in a First order formalism}
  Class.\ Quant.\ Grav.\  {\bf 25}, 095020 (2008)
  [arXiv:0802.2527 [gr-qc]]. 
  
 \bibitem{Corichi:2010ur}
  A.~Corichi and E.~Wilson-Ewing,
  \emph{Surface terms, Asymptotics and Thermodynamics of the Holst Action},
  Class.\ Quant.\ Grav.\  {\bf 27}, 205015 (2010)
  [arXiv:1005.3298 [gr-qc]].
  
 % \bibitem{Liko:2008rr} 
  %T.~Liko and D.~Sloan,
  %\emph{First-order action and Euclidean quantum gravity}
  %Class.\ Quant.\ Grav.\  {\bf 26}, 145004 (2009)
  %[arXiv:0810.0297 [gr-qc]].
    
  \bibitem{Rovelli:1997na} 
  C.~Rovelli and T.~Thiemann,
  \emph{The Immirzi parameter in quantum general relativity},
  Phys.\ Rev.\ D {\bf 57}, 1009 (1998)
  [gr-qc/9705059].
  
  \bibitem{Ashtekar:1997yu} 
  A.~Ashtekar, J.~Baez, A.~Corichi and K.~Krasnov,
  \emph{Quantum geometry and black hole entropy},
  Phys.\ Rev.\ Lett.\  {\bf 80}, 904 (1998)
  [gr-qc/9710007].
  
  
  \bibitem{Ashtekar:1999wa} 
  A.~Ashtekar, A.~Corichi and K.~Krasnov,
  \emph{Isolated horizons: The Classical phase space},
  Adv.\ Theor.\ Math.\ Phys.\  {\bf 3}, 419 (1999)
  [gr-qc/9905089].
  
  \bibitem{Ashtekar:1998sp}
  A.~Ashtekar, C.~Beetle and S.~Fairhurst,
  \emph{Isolated horizons: A Generalization of black hole mechanics}
  Class.\ Quant.\ Grav.\  {\bf 16}, L1 (1999)
  [gr-qc/9812065].
  
  
  
  \bibitem{Ashtekar:1999yj}
A.~Ashtekar, C.~Beetle and S.~Fairhurst,
  \emph{Mechanics of isolated horizons}
  Class.\ Quant.\ Grav.\  {\bf 17}, 253 (2000)
  [gr-qc/9907068].
  
  
  \bibitem{Chatterjee:2006vy}
  A.~Chatterjee and A.~Ghosh,
  \emph{Generic weak isolated horizons}
  Class.\ Quant.\ Grav.\  {\bf 23}, 7521 (2006)
  [gr-qc/0603023].
  
  
  \bibitem{Chatterjee:2008if}
  A.~Chatterjee and A.~Ghosh,
  \emph{Laws of Black Hole Mechanics from Holst Action},
  Phys.\ Rev.\ D {\bf 80}, 064036 (2009)
  [arXiv:0812.2121 [gr-qc]].
  
  
  \bibitem{Durka:2011yv} 
  R.~Durka and J.~Kowalski-Glikman,
  \emph{Gravity as a constrained BF theory: Noether charges and Immirzi 
parameter},
  Phys.\ Rev.\ D {\bf 83}, 124011 (2011)
  [arXiv:1103.2971 [gr-qc]].
  
    R.~Durka,
  \emph{Immirzi parameter and Noether charges in first order gravity},
  J.\ Phys.\ Conf.\ Ser.\  {\bf 343}, 012032 (2012)
  [arXiv:1111.0961 [gr-qc]].
  
  
\bibitem{Hajian:2013lna} 
  K.~Hajian, A.~Seraj and M.~M.~Sheikh-Jabbari,
  \emph{NHEG Mechanics: Laws of Near Horizon Extremal Geometry (Thermo)Dynamics}
  JHEP {\bf 1403}, 014 (2014)
  [arXiv:1310.3727 [hep-th], arXiv:1310.3727].

  

  
 
  
  
  \bibitem{silva}
S.~Silva,
  \emph{Black hole entropy and thermodynamics from symmetries},
  Class.\ Quant.\ Grav.\  {\bf 19}, 3947 (2002)
  [hep-th/0204179].
  
  \bibitem{Ghosh:2011fc} 
  A.~Ghosh and A.~Perez,
  \emph{Black hole entropy and isolated horizons thermodynamics},
  Phys.\ Rev.\ Lett.\  {\bf 107}, 241301 (2011)
  [Erratum-ibid.\  {\bf 108}, 169901 (2012)]
  [arXiv:1107.1320 [gr-qc]].
  
  \bibitem{Majhi:2011ws} 
  B.~R.~Majhi and T.~Padmanabhan,
  \emph{Noether Current, Horizon Virasoro Algebra and Entropy},
  Phys.\ Rev.\ D {\bf 85}, 084040 (2012)
  [arXiv:1111.1809 [gr-qc]].
  
  
  S.~-J.~Zhang and B.~Wang,
  \emph{Surface term, Virasoro algebra and Wald entropy of black holes in higher curvature gravity},
  Phys.\ Rev.\ D {\bf 87}, no. 4, 044041 (2013)
  [arXiv:1212.6896 [hep-th]].

\bibitem{Liko:2011cq} 
  T.~Liko,
  \emph{Barbero-Immirzi parameter, manifold invariants and Euclidean path integrals}
  Class.\ Quant.\ Grav.\  {\bf 29}, 095009 (2012)
  [arXiv:1111.6702 [gr-qc]].
  
  

\end{thebibliography}
\end{document}